\documentclass{ws-procs975x65}

\def\beq{\begin{equation}}
\def\eeq{\end{equation}}

\begin{document}

\title{The mass of dark scalar and phase space analysis of realistic models of static spherically symmetric objects }

\author{Fiziev, Plamen Petkov}

\address{Bogoliubov Laboratory of Theoretical Physics, Joint Institute of Nuclear Research, \\
 141980 Dubna, Moscow Region, Russia\\
 and \\
 Sofia University Foundation for Theoretical and Computational Physics and Astrophysics,\\
Blvd 5 James Bourchier, Sofia 1164, Bulgaria\\
E-mail: fiziev@theor.jinr.ru and fiziev@phys.uni-sofia.bg}

\begin{abstract}
We consider singularities of static spherically symmetric objects in minimal dilatonic gravity.
They are only partially studied and purely understood even in the simplest models of extended gravity.
We introduce the proper form of the structure equations and derive a set of all singularities, which turn to form
several types of sub-manifolds of the phase space. We also introduce for the first time the Lyapunov function
for the corresponding system, its equation, and its basic properties.
The dependence on the mass of the dark scalar is discussed.

\end{abstract}


\bodymatter


\section{Introduction: The basic equations and definitions}
We consider singularities of static spherically symmetric objects (SSSO) in minimal dilatonic gravity (MDG).
They are only partially studied and purely understood even in the simplest models of extended gravity,
see \cite{Fiziev15a} and the references therein.

We present a novel approach to these problems
having in mind that the good physics requires adequate mathematical tools.
Our analysis of the singularities in the phase-space of SSSO in MDG yields new physical consequences.
In particular, it raises a nontrivial problem of bifurcations of the phase-space srtucture of solutions.
A basic example is neutron stars in MDG \cite{Fiziev15b,Fiziev14a,Fiziev14b}.
Our consideration is much more general and can be applied
to any single SSSO in the Universe with metric
$ds^2=e^{\nu(r)}(cdt)^2-dr^2/\left(1-{\frac \varkappa {4\pi r}}\left(m(r)+{\frac 4 3}\pi r^3\varrho\right)\right)-r^2d\Omega^2$,
$r$ being the luminosity radius..

The ODEs for the structure of MDG-SSSO
and the corresponding boundary conditions at the center, at the edge of the SSSO,
and at the boundary of the MDG Universe were derived in \cite{Fiziev15a}
and used in \cite{Fiziev15a,Fiziev15b,Fiziev14a,Fiziev14b}
for neutron stars with different matter equation of state.
One has to solve the non-autonomous system
\begin{eqnarray}
{\frac {dm}{dr}}&=&4\pi r^2\rho_{eff}/\Phi,\quad
{\frac {dp}{dr}}=- {\frac \varkappa {8\pi}}\left(p+\rho c^2\right){\frac{\beta}{r\Delta_\Phi} }, \quad
{\frac {d\Phi}{dr}}=-\varkappa\,{\frac { r^2 p_{{}_\Phi}c^{-2}} {2\Delta}},\label{5ODE}\\
{\frac {dp_{{}_\Phi}}{dr}}&=&- {\frac{ p_{{}_\Phi}}{r\Delta}}\left(3r -{\frac {7\varkappa} {8\pi}} m-{\frac {2\varkappa} 3}\varrho r^3+
{\frac \varkappa 2} r^3\rho_{eff}/\Phi\right)
-{\frac{2}{r}}\rho_{{}_\Phi}c^2, \quad
{\frac {d\nu}{dr}}={\frac \varkappa {4\pi}}{\frac{\beta}{r\Delta_\Phi} }, \nonumber
\end{eqnarray}
for five unknown functions $m(r)$, $\rho(r)$, $\Phi(r)$, $p_{{}_\Phi}(r)$, and $\nu(r)$.

In the Eqs. \eqref{5ODE}
$\Delta=r-{\frac \varkappa {4\pi }}\left(m(r)+{\frac {4\pi} 3} r^3\varrho\right)$,
$\Delta_\Phi=\Phi\Delta-{\frac \varkappa 4}r^3 p_\Phi c^{-2}$,
$\beta=m\Phi+4\pi r^3 p_{eff}c^{-2}$, $p=p(\rho)$ defines the MEOS, $\rho_{eff}=\rho+\rho_\Lambda+\rho_\Phi$,
$p_{eff}=p+p_\Lambda+p_\Phi$, $\rho_\Lambda=\varrho(U(\Phi)-\Phi)$, $p_\Lambda=-\varrho c^2(U(\Phi)-\Phi/3)$,
$\rho_\Phi=p c^{-2}-\rho/3-\varrho V_{,\Phi}+{\frac \varkappa {16\pi}{\frac {\beta p_\Phi c^{-2}}{\Delta_\Phi}}}$.
The comma denotes differentiation with respect to variable $\Phi$.

The three physical parameters in Eqs. \eqref{5ODE}:
$\varkappa \approx 1.8663\times 10^{-27}\,cm/g$ (Einstein constant), cosmological density
$\varrho=\Lambda/\varkappa \approx 5.83 \times 10^{-30}\,g/cm^{3}$, and the velocity of light $c$
are known with different precision,
$\Lambda\approx 1.0876\times10^{-56}\, cm^{-2}$
being the observed value of the cosmological constant.

The two related potentials
\begin{eqnarray}
V(\Phi;\mu)&=&{\frac {\mu^2} {2}}\left(\Phi+1/\Phi-2\right)=\mu^2\big(\text{Z}(\Phi)-1\big)\quad - \text{the dilatonic potential},\label{VU}\\
U(\Phi;\mu)&=&\Phi^2\left(1+{\frac {3\mu^2} {16}}\left(1-1/\Phi^2\right)^2\right)\quad - \text{the cosmological potential},\nonumber
\end{eqnarray}
define the simplest one-parameter-family of withholding potentials \cite{Fiziev13}.
Here we use the Zhukovsky function $\text{Z}(\Phi)={\frac 1 2}(\Phi+1/\Phi)$
in the open interval $\Phi \in (0,\infty)$. To some extent, in the problems under consideration
the potentials \eqref{VU} play the role of the familiar harmonic oscillator potentials in mechanics.
In cosmological units: $\Lambda=1$, $\varkappa=1$, $c=1$ the dimensionless parameter
$\mu ={\frac {m_\Phi c}{\hbar\sqrt{\Lambda}}} \sim 10^{3} \div 10^{34}$,
being largely unknown, defines the mass of the dilaton $m_\Phi=\mu \hbar\sqrt{\Lambda}/c \sim 10^{-3} \div 10^{28}\, eV/c^2$.

The generalization to the case of potentials with $n$-in-number minima
depends on $(n+1)$-in-number dimensionless parameters $\mu,z_1,\dots,z_n$ and reads
\beq
V(\Phi;\mu,z_1,\dots,z_n)=\mu^2\prod_{k=1}^n\big(\text{Z}(\Phi)-z_k\big).
\label{VU_n}
\eeq

The introduced and utilized in \cite{Fiziev15a} dark scalar $\varphi=\left(\ln\left(1+\ln(\Phi)\right)\right)^{1/n},\,\,n=1,3,5;$
has the same mass $m_\Phi$ by construction and stretches the physical domain of the dilaton.
This facilitated studies of the very challenging numerical problems under consideration
and made it possible to discover new phenomena\cite{Fiziev15a}.

In Fig.~1, we sketch the picture of the MDG Universe with a single SSSO in it.
\begin{figure}[ht]
\begin{center}
\vskip .in
\includegraphics[width=3.2in]{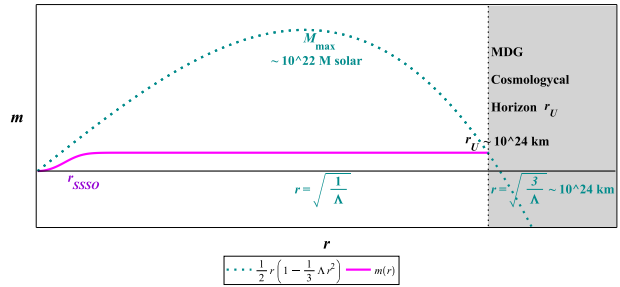}
\end{center}
\vskip .in
\caption{MDG (the Klotter-Weyl like) Universe  with a single SSSO in it. As seen, there exist three very different scales: the size of the SSSO
$r_{SSSO} \sim  10^{5}\div 10^{20}$ cm, the size of the Universe $r_U \sim 10^{28}$ cm,
and the completely unknown Compton length of the dilaton $\lambda_\Phi \sim 10^{-3} \div 10^{28}$ cm.
This makes the numerical studies of the problem an extremely difficult issue .}
\label{Fig1}
\end{figure}

\section{The Phase space of the problem and its vector field}

We rewrite the system \eqref{5ODE} as an autonomous system of six ODEs
\beq
\dot r=F_r,\quad\dot m=F_m,\quad\dot p=F_p,\quad\dot \Phi=F_\Phi,\quad\dot p_\Phi=F_{p_\Phi},\quad\dot \nu=F_\nu
\label{6ODE}
\eeq
in the 6d phase space $\mathbb{M}^{(6)}_{r,m,p,\Phi,p_\Phi,\nu}$ introducing
the regularization parameter $\tau$: $dr=r\Delta\Delta_\Phi d\tau$
and denoting by dot the differentiation with respect to it.

The six functions $F_{\!...}$ in Eqs. \eqref{6ODE} are of the type $F_{\!...}(r,m,p,\Phi,p_\Phi;\mu)$:
\begin{eqnarray}
F_r&=&r\Delta\Delta_\Phi,\,\,
F_m={\frac {4\pi r^3}{\Phi}}\left(\alpha\Delta_\Phi+{\frac{1}{16\pi}}p_\Phi\beta \right)\Delta,
\,\, F_p=-{\frac{1}{8\pi}} \left(\rho+p\right)\beta\Delta,\label{F} \\
F_\Phi&=&{\frac{1}{2}}p_\Phi\Delta_\Phi,\,\,
F_{p_\Phi}={\frac{1}{\Phi}}\left(\alpha_1\Delta+\alpha_2\Delta_\Phi+2\alpha_3\Delta\Delta_\Phi+
{\frac{1}{32\pi}}r^3p_\Phi\beta\right),\,\, F_\nu={\frac{1}{4\pi}} \beta\Delta,\nonumber
\end{eqnarray}
where now $\Delta=r-{\frac 1 {4\pi }}\left(m(r)+{\frac {4\pi} 3} r^3\right)$,
$\Delta_\Phi=\Phi\Delta-{\frac 1 4}r^3 p_\Phi$,
$\beta=m\Phi+4\pi r^3 p_{eff}$,
$\rho_\Lambda=U(\Phi)-\Phi$,
$p_\Lambda=-(U(\Phi)-\Phi/3)$,
$\rho_\Phi=-T/3- V_{,\Phi}+{\frac 1 {16\pi}{\frac {\beta p_\Phi}{\Delta_\Phi}}}$, $T=\rho-3p$,
and
\begin{eqnarray}
\alpha &=& p+{\frac 2 3}\left(V_{,\Phi}+U-\Phi\right),\,\,
\alpha_1={\frac{1}{8\pi}}\Phi p_\Phi\beta,\nonumber\\
\alpha_2 &=&\Phi p_\Phi\left(3r-{\frac{1}{2\pi}},
\left({\frac 7 4}m+{\frac{4}{3}}\pi r^3\right)+{\frac 1{2\Phi}}r^3\alpha\right),\,\,
\alpha_3=\Phi V_{,\Phi}-{\frac 1 3}T .
\label{alpha}
\end{eqnarray}

Since the functions \eqref{F}  do not depend on the variable $\nu$, the last of the Eqs. \eqref{6ODE} splits out
and we can consider the reduced (sub)system of order five in the phase space $\mathbb{M}^{(5)}_{r,m,p,\Phi,p_\Phi}$.
Hence, after all we have to consider a 5d vector field
${\bf F}=\{F_r,F_m,F_p,F_\Phi,F_{p_\Phi}\} \in T \mathbb{M}^{(5)}_{r,m,p,\Phi,p_\Phi}$.
It depends on only one parameter $\mu$ in the case of potentials \eqref{VU}, when one uses cosmological units.

\section{Singular submanifolds of the phase space $\mathbb{M}^{(5)}_{r,m,p,\Phi,p_\Phi}$}

A specific peculiarity of our problem is that
its singular points are not isolated and form singular sub-manifolds
${}_iS\mathbb{M}^{(d_i)}\subset \mathbb{M}^{(5)}_{r,m,p,\Phi,p_\Phi}$
of different dimensions $d_i$.
Therefore, to be able to consider the
changes of structure of the phase-space flow of SSSO in MDG,
we are forced to develop the very theory of bifurcations.
In the present article, we make some first steps toward the solution of this problem.

The equilibria, dubbed also singular points of Eqs. \eqref{6ODE} in the phase space $\mathbb{M}^{(5)}_{r,m,\rho,\Phi,p_\Phi}$,
are zeros of the system ${\bf F}=0$.
One has to impose also some additional restrictions on the physically admissible equilibria.
As seen from the first of Eqs. \eqref{F}, there exist only three types of equilibrium points.

1) The center of SSSO: $r=0$ $\Rightarrow$ $F_r=0$, $F_m=0$, and $F_\Phi=0$.
We equate $m=0$ and $p_\Phi={\frac 2 9}T-{\frac 2 3} V_{\!{,\Phi}}$
to nullify the rest of the components of the vector ${\bf F}$
having additional physical constraint: {\it a finite\/} value $p_\Phi|_{r=0}$  \cite{Fiziev15a}.
These conditions define a two-dimensional singular manifold ${{}_{{}_1}}S\mathbb{M}^{(2)}_{p,\Phi}$
with inner coordinates ${p,\Phi}$.

2) The cosmological horizon:  $\Delta=0$ $\Rightarrow$ $F_r=0$, $F_m=0$, and $F_p=0$.
We equate $p_\Phi=0$ and impose the physical requirement $\Phi=1$
to nullify the rest of the components of the vector ${\bf F}$ having {\it the de Sitter\/} vacuum $\Phi|_{r=r_U}=1$
\cite{Fiziev15a}.

There exist two different cases:
a) Inside the SSSO ($r\leq r_{\!{}_{SC}}, \rho>0, p>0$) $\Rightarrow$  ${{}_{{}_2}}S\mathbb{M}^{(2)}_{r,p}$.
b) Outside the SSSO ($r > r_{\!{}_{SC}},\rho=0, p=0$) $\Rightarrow$  ${{}_{{}_3}}S\mathbb{M}^{(1)}_{\,\,r}$.
\begin{figure}[ht]
\vskip -.in
\begin{center}
\includegraphics[width=4in]{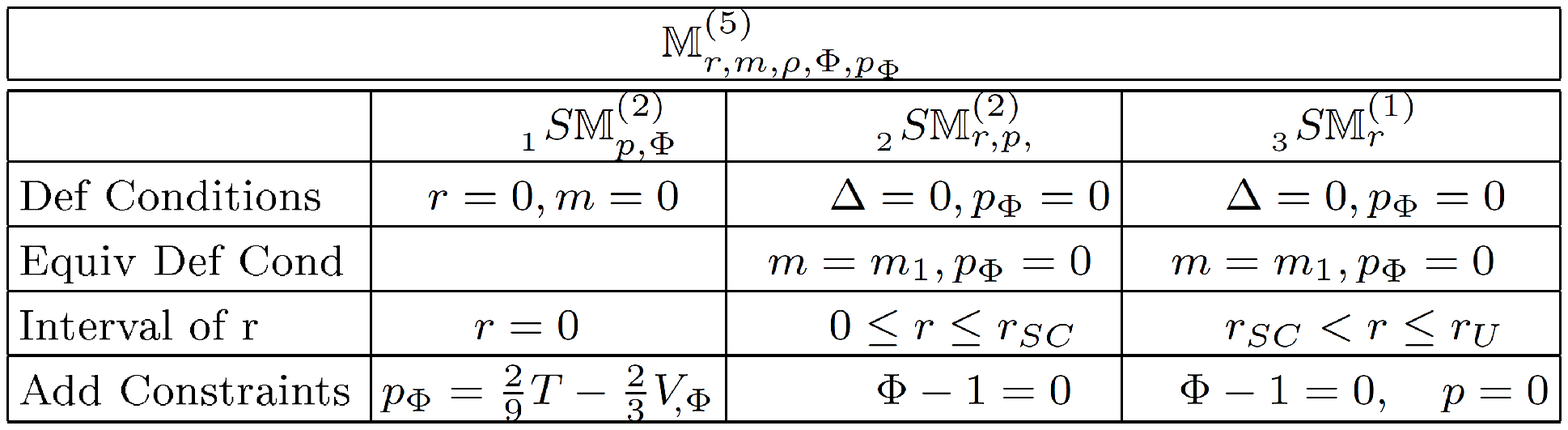}
\end{center}
\vskip .in
\caption{The singular manifolds ${{}_{{}_{1,2,3}}}S\mathbb{M}^{(d_{1,2,3})}_{\dots}\subset \mathbb{M}^{(5)}_{r,m,p,\Phi,p_\Phi}$.}
\label{FigT1}
\end{figure}

3) The  singular manifold, defined by  $\Delta_\Phi=0$ $\Rightarrow$  $F_r=0$, $F_\Phi=0$.
We have to equate $\beta=0$ to nullify the rest of the components of the vector ${\bf F}$.

There exist two different cases:
a) Inside the SSSO ($r\leq r_{\!{}_{SC}}, \rho>0, p>0$)  $\Rightarrow$
${{}_{{}_4}}S\mathbb{M}^{(3)}_{r,p,\Phi}$.
b) Outside the SSSO ($r > r_{\!{}_{SC}},\rho=0, p=0$)  $\Rightarrow$
${{}_{{}_5}}S\mathbb{M}^{(2)}_{r,\Phi}$.
\begin{figure}[ht]
\vskip -.in
\begin{center}
\includegraphics[width=4in]{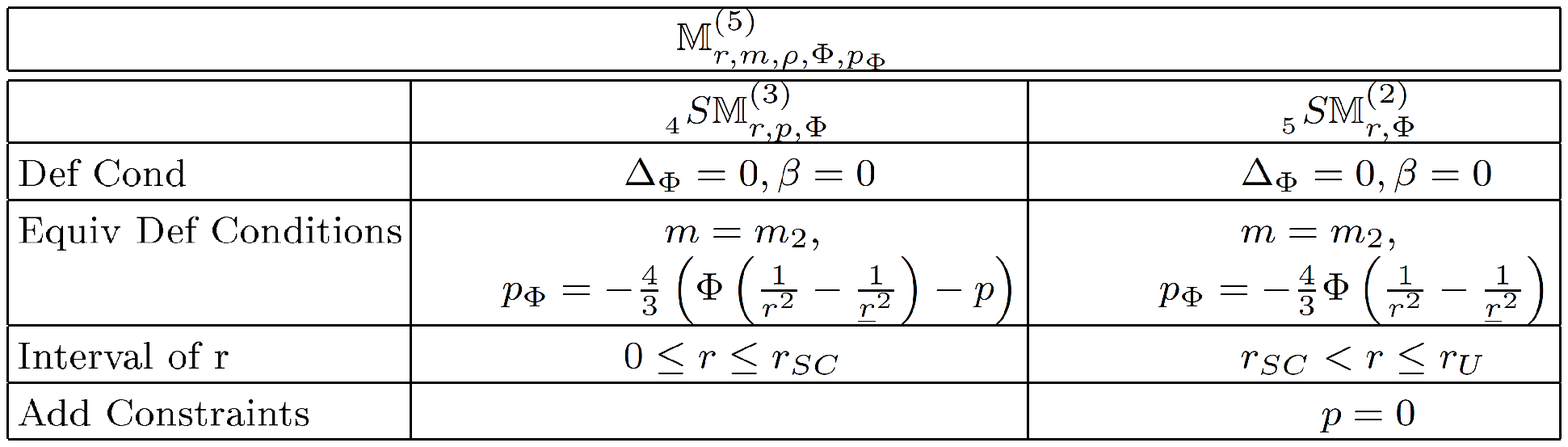}
\end{center}
\vskip .in
\caption{The singular manifolds ${{}_{{}_{4,5}}}S\mathbb{M}^{(d_{4,5})}_{\dots}\subset \mathbb{M}^{(5)}_{r,m,p,\Phi,p_\Phi}$.}
\label{FigT2}
\end{figure}
Besides the singular manifolds ${}_{1\dots 5}S\mathbb{M}$ there are no more physical singular points\footnote{The
singularities for values $\Phi=0$ and $\Phi=\infty$ are excluded by the withholding condition \cite{Fiziev13}.}.
Contrary to the standard situation, there do not exist any isolated singular points.

\begin{figure}[ht]
\begin{center}
\vskip -.in
\includegraphics[width=3.2in]{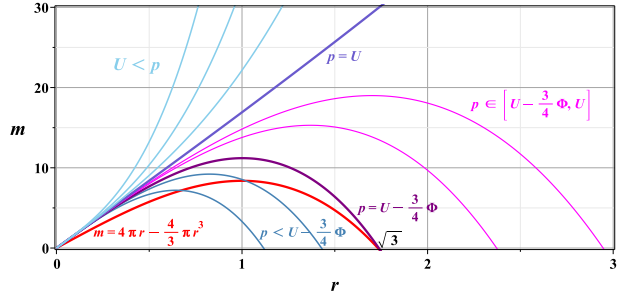}
\end{center}
\vskip .in
\caption{The complicated picture of projections of some singular manifolds on the $m-r$ plane :
Red line presents the function $m=4\pi r-{\tfrac 4 3}\pi r^3$; Other lines in different colors present
the function $m=4\pi r-{\tfrac 4 3}\pi r^3+{\frac 4 3}\pi r^3\left({\frac 1 {r^2}}-{\frac 1 {\underline{r}^2}}\right)$
for pointed in the figure values of $p>0$, and  fixed value of $\Phi$.}
\label{Fig4}
\end{figure}

The final results are presented in Figs.~\ref{FigT1},\ref{FigT2}. We also use the quantities
$\underline{r}=\underline{r}(\Phi,p;\mu)=\sqrt{{{\Phi}/{ U(\Phi;\mu)-p}}}$,
$m_1=4\pi r-{\tfrac 4 3}\pi r^3$,
and $m_2=4\pi r-{\tfrac 4 3}\pi r^3+{\frac 4 3}\pi r^3\left({\frac 1 {r^2}}-{\frac 1 {\underline{r}^2}}\right)$
(in cosmological units)
to illustrate different positions of singular manifolds for different values of the parameters, see Fig.~\ref{Fig4}.
The change of these positions with change of the parameters leads to bifurcations of the phase portrait.
This issue needs a more detailed study which will be presented elsewhere.

\section{The Lyapunov function for SSSO structure equations}

The power of the Lyapunov analysis
of the MDG-Friedmann-Robertson-Walker time-dependent model of Universe
was demonstrated for the first time in \cite{Fiziev02,Fiziev03}.
Since in the case of SSSO we consider the dependence
of solutions on the luminosity radius $r$,
taking into account the metric signature we obtain a similar problem with
the dilatonic potential $V$ replaced by $-V$, see, for example,
\cite{Frolov08,Sotiriou10} and the references therein.
The analysis of of solutions
based on nonexisting pseudo-energy-conservation
was presented in the framework the $f(R)$-models \cite{Frolov08,Sotiriou10}.
This issue needs a careful reexamination using the Lyapunov analysis.

For SSSO  the dilaton equation $\Box\Phi+  V_{\!{}_{,\Phi}}(\Phi) = {\frac 1 3} T$
takes the form of Eq.(A.1d) from paper \cite{Fiziev15a}.
It can be rewritten in the form of the Lyapunov equation
\beq
{\frac {d\eta}{dl}}=W(r,m,p,\Phi,p_\Phi;\mu),
\label{deta}
\eeq
for the Lyapunov function
\beq
\eta ={\frac 1 2} \left( { \frac {d\Phi}{dl} } \right)^2- V(\Phi;\mu)={\frac 1 2} v^2- V(\Phi;\mu).
\label{eta}
\eeq
\begin{figure}[ht]
\begin{center}
\vskip -.in
\includegraphics[width=4.8in]{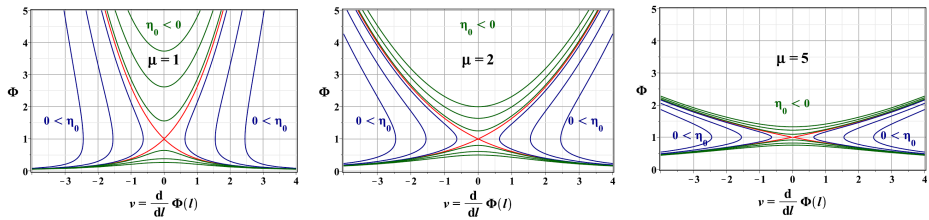}
\end{center}
\vskip .in
\caption{The level-curves of the function \eqref{eta} and their dependence of the mass-parameter $\mu$.
As seen, the increase of $\mu$ drastically deforms and shrinks the level's picture.}
\label{Fig5}
\end{figure}
The level-curves of this function and their dependence of the mass-parameter $\mu$ are shown in Fig.~5.
We use the variable $v={\frac{d\Phi}{dl}}$, the potential $V(\Phi)$ \eqref{VU},
and the true geometrical distance $dl=r(\tau)\Delta_\Phi(\tau)\sqrt{r(\tau)\Delta(\tau)}d\tau$.
In Eq. \eqref{deta}
\beq
W=\left({\frac 1 3}T+{\frac 3 2} {\frac {\Phi}{r\Delta_\Phi}}{\frac {m-m_2}{4\pi}}v\right)v,
\,\,\, \text{where}\,\,\,v=-{\frac 1 2}{\frac {r^{3/2}}{\Delta^{1/2}}}\,p_\Phi.
\label{W}
\eeq
The sign of this function defines the direction of change of the Lyapunov function $\eta$.
This direction may be quite counterintuitive in some domains of the phase-space.
We intend to consider the corresponding important results somewhere else.

Here we would like to mention that, as shown in \cite{Fiziev13}, MDG is, in general, only locally equivalent to the $f(R)$ extended theories of gravity,
see for example \cite{Frolov08,Sotiriou10,Starobinsky80,Felice10,Capozziello11,Nojiri11,Capozziello11a}, and a large amount of references therein.
As a rule, the presented results can be applied to the $f(R)$ theories only to some extent and such application requires special cares.
The exception are the $f(R)$ theories which are globally equivalent to MDG, but one is not able to find these models in the existing literature.

\section*{Acknowledgments}
The author is grateful to S. Capozziello, M. De Laurentis, and S. Odintsov for very useful discussions on the subject of the present paper.
This research was supported in part by the Sofia University Foundation for Theoretical and Computational Physics and Astrophysics
and by the Bulgarian Nuclear Regulatory Agency, Grant for 2015, as well as by “NewCompStar,” COST Action
MP1304.

\end{document}